\documentclass[aps,prb,showpacs,twocolumn,superscriptaddress]{revtex4}
\usepackage{graphicx}

\begin{document}
\title{Coulomb
Interaction Effects on the Terahertz Photon-Assisted Tunneling
through a InAs Quantum Dot}
\author{R.-Y. Yuan}
\affiliation{Center for Theoretical Physics, Department of Physics,
Capital Normal University, Beijing 100048, China}
\author{G.-B. Zhu}
\affiliation{Department of Physics, Heze University, Heze, Shandong
274015, China}
\author{X. Zhao}
\affiliation{Department of Physics, Capital Normal University,
Beijing 100048, China}
\author{Y. Guo}
\affiliation{Department of Physics and State Key Laboratory of
Low-Dimensional Quantum Physics, Tsinghua University, Beijing
100084, China} \affiliation{Collaborative Innovation of Quantum
Matter, Beijing, China}
\author{H. Yan}
\affiliation{Laboratory of Thin Film Materials, Beijing University
of Technology, Beijing 100022, China}
\author{Q. Sun}
\email{sunqing@iphy.ac.cn} \affiliation{Center for Theoretical
Physics, Department of Physics, Capital Normal University, Beijing
100048, China}
\author{A.-C. Ji}
\email{andrewjee@sina.com} \affiliation{Center for Theoretical
Physics, Department of Physics, Capital Normal University, Beijing
100048, China}

\date{{\small \today}}

\begin{abstract}
Recently, the terahertz (THz) photon-assisted tunneling (PAT)
through a two-level InAs quantum dot (QD) has been successfully
realized in experiment [Phys. Rev. Lett. {\bf 109}, 077401 (2012)].
The  Coulomb interaction in this device is comparable with the
energy difference between the two energy levels. We theoretically
explore the effects of Coulomb interaction on the PAT processes and
show that the main peaks of the experiment can be well derived by
our model analysis. Furthermore, we find additional peaks, which
were not addressed in the InAs QD experiment and may be further
identified in experiments. In particular, we show that, to observe
the interesting photon-induced excited state resonance in InAs QD,
the Coulomb interaction should be larger than THz photon frequency.
\end{abstract}


\maketitle

\section {Introduction}
In various nano-structures, such as a single electron transistor
\cite{Kouwenhoven,Oosterkamp,Suzuki}, applying a time varying
oscillating potential to the Coulomb island can induce an inelastic
tunneling event known as photon-assisted tunneling (PAT). PAT has
been intensively studied, because it can be exploited to build a
highly sensitive detector \cite {Tucker} or a solid-state quantum
bit \cite{Yamamoto,Hayashi,Petta,Koppens}. So far, these studies
were mainly performed in the GHz range, namely microwave-field (MWF)
\cite{Platero,Goker,Braakman,Bergenfeld,Frey,Wiel,Taranko}. On the
other hand, the terahertz (THz) region, which is of fundamental
industrial importance, has attracted broad scientific interest in
the past decade \cite{Ferguson,Tonouchi}. However, the THz devices,
especially in the region of 1-10 THz, have not been fully developed,
giving rise to the so-called ``THz gap".

Recently, experiments have been performed to implement a THz
detector via the THz photon-assisted tunneling through a carbon
nanotube \cite{Fuse,Kawano}  or  self-assembled InAs quantum dot
(QD) \cite{Jung,Hirakawa}, where it was shown  that both the
charging and orbital quantization energies are typically 10-40 meV,
which correspond to the THz region (2.5-10 THz). In the early works
by the group of Ishibashi \cite{Fuse,Kawano}, the authors found that
the carbon nanotubes QD can lead to new side peaks that appear only
under the THz irradiations. Very recently, the THz PAT in a single
self-assembled two-level InAs QD with $s$ and $p$ orbitals, has also
been investigated by Shibata {\it et al.} \cite{Shibata}. They
showed that, in addition to the PAT processes of  $s$ level and the
Coulomb blockade oscillation of $p$  level, an interesting
photon-induced excited state resonance (PIER) of  $p$  level can be
observed, when THz photon frequency is larger than the energy
difference.

The irradiation induced PAT side peaks and the PIER  in the presence
of Coulomb blockade  were first  investigated in the MWF devices.
Note that in a two-level MWF QD, the Coulomb interaction is much
larger than the energy difference and regarded as infinite, all the
Coulomb interaction related resonances are ignored
\cite{Oosterkamp}. Whereas in a InAs QD, because the energy
difference lies in the THz region, the intra-dot Coulomb interaction
becomes comparable with the energy difference. In this case, both
the PAT and Coulomb blockade effects are involved together, and the
finite Coulomb interaction may present new features on the resonant
tunnelings beyond the MWF QD.

\begin{figure}[h]
\includegraphics[width=0.5\textwidth]{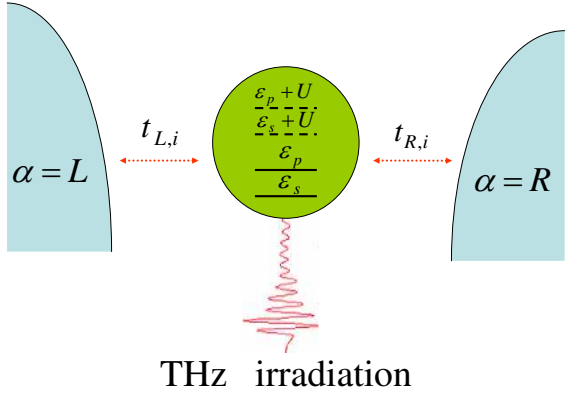}
\caption{\label{Fig1} (color online)  Schematic diagram for the
system of a two-level QD under the THz irradiation with frequency
$\omega$. Two reservoirs of 2DEG are connected with the center
region, where $\varepsilon_{i=s,p}$ denotes the $s$ and $p$ energy
levels in InAs QD. $t_{\alpha, i}$ describe the hopping matrix
elements between leads and the two energy levels in the center
region. $U$ is the intra-dot Coulomb interaction and
$\varepsilon_i+U$ are the Coulomb interaction related  energy
levels.
 \label{fig1}}
\end{figure}

In this paper, we theoretically explore the Coulomb interaction
effects on the PAT processes in InAs QD, see  Fig. \ref{fig1} for
the schematic diagram. We demonstrate that, the presented peaks of
our model analysis $\varepsilon_{s}$, $\varepsilon_p + U$,
$\varepsilon_s \pm\omega$, $\varepsilon_{p}$ agree well with $E_0$,
$E_1$, $E_0\pm hf_{\rm THz}$, and the PIER of $p$ level  in the
experiment \cite{Shibata}. On the other hand, we find  the side peak
$\varepsilon_p-\omega$ induced by the THz irradiation in our  model
analysis. This peak can be identified in the original experimental
data, but was not addressed in  the reference \cite{Shibata}. In
addition, beyond the Coulomb blockade oscillation peak
$\varepsilon_p + U$, we also find there exists the $\varepsilon_s +
U$ peak. This peak seems not readily discriminated from $E_1$
($E_1\equiv\varepsilon_p + U$) peak in the experiment. One may
expect to identify both the peaks $\varepsilon_{s,p} + U$, which
were not observed in the MWF QD due to infinite Coulomb interaction,
by increasing the separation between the energies $\varepsilon_s$
and $\varepsilon_p$ in future experiment.
In particular, we show
that, to observe the interesting photon-induced excited state
resonance of $p$ level, the Coulomb interaction should be larger
than  THz photon frequency. These features may be beneficial for
future  THz devices, such as an ultra-sensitive THz detector, which
may also open new possibilities of controlling carrier dynamics in
quantum nanostructures by THz radiation.

The organization of this paper is as follows:  In Sec. II, we first
present the tunneling model of the two-level InAs QD under THz
irradiation. Then, we derive the formulation of average currents in
Sec. III. In Sec. IV, we  analyze the InAs QD experiment and explore
the Coulomb interaction effects on the resonance tunnelings. Sec. V
is the conclusion.

\section {Model of The THz PATs
through a two-level QD} We consider a system of a two-level
tunneling QD under the THz irradiation as depicted  in Fig.
\ref{fig1}, where $\varepsilon_{i=s,p}$ denotes the $s$ and $p$
energy levels in InAs QD. The dot is connected to two electronic
reservoirs with chemical potentials $\mu_\alpha$, $\alpha=L,R$ and
$\mu_L-\mu_R=eV$. Then, the Hamiltonian of the system can be
described by
\begin{equation}
H(t)=H_{\rm lead}+H_{\rm d}(t)+H_{\rm t},\label{Hamiltonian}
\end{equation}
where the first term $H_{\rm lead}=\sum_{\alpha}
\varepsilon_{\alpha,k}c^{\dagger}_{\alpha,k}c_{\alpha,k}$ describes
the left and right leads respectively. $\varepsilon_{\alpha,k}$ is
the single-electron energy, and $c^{\dagger}_{\alpha,k}$
($c_{\alpha,k}$) is the creation (annihilation) operator of the
electrons in leads. The second term denotes the Hamiltonian of the
central InAs QD, where we have taken into account the THz
irradiation and intra-dot Coulomb interaction, which is given by
\begin{equation}
H_{\rm
d}(t)=\sum_{i=s,p}\varepsilon_{d,i}(t)d^{\dagger}_{i}d_{i}+\frac{U}{2}
n_{i}n_{\bar{i}},
\end{equation}
with $\varepsilon_{d,i}(t)=\varepsilon_{i}+W_{d}(t)$ denoting the
time-dependent energy levels of the QD under the THz fields. Here,
we have implemented the adiabatic approximation by introducing the
THz irradiation as an oscillating potential  with
$W_{d}(t)=W_{d}\cos\omega t$ \cite{Wingreen,Jauho}, which causes a
rigid shift of  $\varepsilon_{i}$. $d^{\dagger}_{i}$ ($d_{i}$) is
the creation (annihilation) operator in the QD and $U$ represents
the intra-dot Coulomb repulsion between the $s$ and $p$ energy
levels with $i$ ($\bar{i}$)$=s$ ($p$) or $p$ ($s$). Finally, the
third term $H_{\rm t}$ describes the tunneling part, which can be
written as
\begin{equation}
H_{\rm
t}=\sum_{\alpha;i=s,p}t_{\alpha,i}c^{\dagger}_{\alpha,k}d_{i}+{\rm
H.c.}.
\end{equation}
Here $t_{\alpha, i}$ are the hopping matrix elements between leads
and two energy levels in InAs QD.

\section {Formulation of the average currents}
Now, we implement the Keldysh non-equilibrium Green's function
method to solve this model, which offers an efficient way to deal
with many-body correlations  in a unified fashion
\cite{Wingreen,Jauho}. By applying this method to the irradiation
problem, the time-dependent current from the left lead to the QD can
be calculated from the evolution of the total number operator of the
electrons in the left lead,
\begin{eqnarray}
I_{L}(t)&\equiv&-e\langle\dot{n}_L\rangle=ie\langle[n_L,H(t)]\rangle
\end{eqnarray}
where $n_L=\sum_{k}c^{\dagger}_{L,k}c_{L,k}$ is the number operator
of the electrons in the left lead.  The total current can be written
as a summation of the time-dependent left-going current through the
$s,p$ level respectively, i.e. $I_{L}(t)=\sum_{i=s,p}I^{i}_{L}(t)$,
which is given by
\begin{equation}
I^{i}_{L}(t)=2e{\rm Re}[t_{L,i}G^{<}_{i,L}(t,t)].
\end{equation}
Here, $G^{<}_{i,L}(t,t)\equiv i \langle c^{\dagger}_{L,k}(t)d_i(t)
\rangle$ is the Keldysh Green function. With the help of the Dyson
equation,  $G^{<}_{i,L}(t,t)$ can be written as
\begin{eqnarray}
 \nonumber G^{<}_{i,L}(t,t)=\int {\rm
d}t_1
t^{*}_{L,i}[G^{r}_{i,i}(t,t_1)g^{<}_{kL}(t_1,t)\\
+G^{<}_{i,i}(t,t_1)g^{r}_{kL}(t_1,t)],
\end{eqnarray}
where $G^{r}_{i,i}(t,t_1)\equiv -i\theta(t-t_1)\langle
\{d_i(t),d^{\dagger}_i(t_1)\} \rangle$, $G^{<}_{i,i}(t,t_1)\equiv i
\langle d^{\dagger}_i(t_1)d_i(t) \rangle$, and $g^{<}_{kL}=i
f_L(\varepsilon-\mu_L)$, $g^{r}_{kL}=-i \theta (t-t_1)$ denote the
less and retarded Green functions of the electron in the left lead.
Taking Eq. (6) into Eq. (5), we arrive at
\begin{eqnarray}
\nonumber I^{i}_{L}(t)= -2e  {\rm Im} \int_{-\infty}^{t}{\rm
d}t_{1}\int \frac{{\rm d}\varepsilon}{2 \pi} e^{-i\varepsilon
(t-t_1)}\\ \Gamma_i^{L}{(\varepsilon)}  [G^{<}_{i,i}(t,t_1)
+f_L(\varepsilon)G^{r}_{i,i}(t,t_1)].
\end{eqnarray}

Next, using the Keldysh equation $G^{<}_{i,i}(t,t^{\prime})=\int
\int{\rm d}t_1 {\rm d}t_2 G^{r}_{i,i}(t,t_1)\Sigma
^{<}_{i,i}(t_1,t_2) G^{a}_{i,i}(t_2,t^{\prime})$ with
$G^{r}_{i,i}=(G^{a}_{i,i})^{*}$, the self-energy function $\Sigma
^{<}_{i,i}(t_1,t_2)=i\int\frac{{\rm d}\varepsilon}{2
\pi}e^{-i\varepsilon(t_1-t_2)}\sum_{\alpha=L,R}f_{\alpha}
(\varepsilon-\mu_{\alpha})\Gamma^{\alpha}_{i}$. Here, under the wide
bandwidth approximation \cite{Wingreen2}, the line width function
$\Gamma^{\alpha}_{i}\equiv
2\pi\sum_{\alpha}t_{\alpha,i}t^{*}_{\alpha,i} \delta
(\varepsilon-\varepsilon_{\alpha,k})$ is independent of energy.

Then, the time-dependent left-going current reduces to
\begin{eqnarray}
I_{L}^i(t)=&-&e\Gamma^{L}_{i}\int\frac{{\rm
d}\varepsilon}{2\pi}{\sum_{\alpha=L,R}f_{\alpha}
(\varepsilon-\mu_{\alpha})|A^{\alpha}_{i}(\varepsilon,t)|^{2}}\nonumber \\
&-&e\Gamma^{L}_{i}\int\frac{{\rm d}\varepsilon}{2\pi}
{2f_{L}(\varepsilon-\mu_{L}){\rm
Im}A^{L}_{i}(\varepsilon,t)},\label{current1}
\end{eqnarray}
where $f_{\alpha} (\varepsilon)=
{1}/({e^{\beta(\varepsilon-\mu_{\alpha})}+1}) $ denotes the Fermi
distribution function of leads, and $A^{\alpha}_{i}(\varepsilon,t)$
is the spectral function which is given by
\begin{equation}
A^{\alpha}_{i}(\varepsilon,t)=\int_{-\infty}^{t}{\rm
d}t_{1}G^{r}_{i,i}(t,t_{1})e^{-i \varepsilon
(t_{1}-t)}.\label{spectrum}
\end{equation}

For the two-terminal device, the average current of each level
$\langle I_i \rangle \equiv \langle I_{L}^i(t) \rangle -\langle
I_{R}^i(t)\rangle$ can be derived with the help of Eq.
(\ref{current1}) as
\begin{eqnarray}
\nonumber \langle I_i \rangle=&-&2e
\frac{\Gamma^{L}_{i}\Gamma^{R}_{i}}{\Gamma^{L}_{i}
+\Gamma^{R}_{i}}\int\frac{{\rm d}\varepsilon}{2\pi}f_{L}
(\varepsilon){\rm Im} \langle
A^{L}_{i}(\varepsilon,t)\rangle\\
&+&2e\frac{\Gamma^{L}_{i}\Gamma^{R}_{i}}{\Gamma^{L}_{i}
+\Gamma^{R}_{i}}\int\frac{{\rm
d}\varepsilon}{2\pi}f_{R}(\varepsilon){\rm Im}\langle
A^{R}_{i}(\varepsilon,t)\rangle,\label{current}
\end{eqnarray}
 and the total average current is  $\langle I \rangle
=\sum_{i\>=s,p}\langle I_{i} \rangle$.

The main step is then to determine the spectral function in Eq.
(\ref{current}). Note that, the spectral function is related to the
retarded Green function $G^{r}_{i,i}$, which can be determined by
the equation of motion (EOM) \cite{Wang,Hackenbroich}. Here, we take
the higher-order approximation to investigate the THz PATs and
obtain
\begin{eqnarray}
\nonumber G^{r}_{i,i}(t,t^{\prime})
=[1-n_{\bar{i}}]g^{r}_{\varepsilon_i}
(t,t^{\prime})e^{-\frac{\Gamma_i}{2}(1-n_{\bar{i}})(t,t^{\prime})}\\
+n_{\bar{i}}g^{r}_{\varepsilon_i+U}
(t,t^{\prime})e^{-\frac{\Gamma_i}{2}n_{\bar{i}}(t,t^{\prime})},\label{retard}
\end{eqnarray}
where $g^{r}_{\varepsilon_i}(t,t^{\prime})\equiv
-i\theta(t-t^{\prime})e^{-i\int^{t}_{t^{\prime}}\varepsilon_i(\tau){\rm
d}\tau}$, and $g^{r}_{\varepsilon_i+U}(t,t^{\prime})\equiv
-i\theta(t-t^{\prime})e^{-i\int^{t}_{t^{\prime}}[\varepsilon_i(\tau)+U]{\rm
d}\tau}$.

In the above equation, there are two kinds of resonances. In the
first term, the resonances are at $\varepsilon_{s(p)}$, which occur
when the $p(s)$ level is empty and there is no  Coulomb interaction
effect between the two levels. Whereas in the second term, the
Coulomb interaction related resonances  are at $\varepsilon_{s(p)} +
U$, which can be understood as follows.  First, in the case where
$p$ level is occupied, if another electron wants to transit through
the $s$ level, due to the Coulomb interaction $U$ between the $s$
and $p$ levels, the energy of this electron becomes $\varepsilon_{s}
+ U$. On the other hand, in the case where $s$ level is occupied, if
another electron wants to transit through the $p$ level, the
resonance  of this electron occurs at $\varepsilon_{p} + U$.

Taking the retarded Green function into the Eq. (9) gives rise to
the following spectral function
\begin{eqnarray}
\nonumber A^{\alpha}_{i}(\varepsilon,t)=\sum_{n}J^{2}_{n}
(\frac{W_{d}}{\omega})e^{in\omega
t}{\frac{1-n_{\bar{i}}}{\varepsilon-\varepsilon_{i}-n\omega+i
\frac{\Gamma_{i}(1-n_{\bar{i}})}{2}}}\\
+\sum_{n}J^{2}_{n}(\frac{W_{d}}{\omega})e^{in\omega
t}{\frac{n_{\bar{i}}}{\varepsilon-\varepsilon_{i}
-U-n\omega+i\frac{\Gamma_{i}n_{\bar{i}}}{2}}},\label{spectrum_1}
\end{eqnarray}
where $J_{n}$ is the $n$-order Bessel function and $n_{i}$ ($n_{\bar
i}$) is the average electron occupation number of each energy level,
which can be derived as
\begin{equation}\label{number}
n_{i}=\langle d^{\dagger}_{i}d_{i}\rangle=\int \frac{{\rm
d}\varepsilon}{2\pi} \sum_{\alpha}f_{\alpha}(\varepsilon)
\Gamma^{\alpha}_{i}\langle|A^{\alpha}_{i}(\varepsilon,t)|^{2}
\rangle.
\end{equation}

Note that, in the two-level MWF QD,  Coulomb interaction is much
larger than the energy difference and can be regarded as infinite.
Therefore, the second term of the spectral function Eq.
(\ref{spectrum_1}) is ignored and only the first term contributes to
the average currents. However, in  InAs QD, because the intra-dot
Coulomb interaction becomes comparable with the energy difference,
the Coulomb blockade oscillations and related PATs arise from the
second term, which is of equal importance as the first term and may
present new features. To our knowledge, the finite $U$ effects have
not been explored theoretically in a InAs QD. Here for simplicity,
we assume $\Gamma^{L}_{i}=\Gamma^{R}_{i}$ and define
$\Gamma_{i}=\Gamma^{L}_{i}+\Gamma^{R}_{i}$. Then, by solving the
Eqs. (\ref{spectrum_1}-\ref{number}) self-consistently, one can
obtain the average currents.

\section{Results of the THz PATs  in the presence of Coulomb
Interaction} In this section, we present and discuss the main
results of this work. We shall first analyze the tunneling
experiment through a InAs QD based on the above formulation. Then,
we explore the Coulomb interaction effects on the resonant
tunnelings. To compare with the experiment, we consider the
following  parameters: $W_d=0.9$, $\Gamma_s=0.01$, $\Gamma_p=0.03$,
$k_BT=0.1$, and $V=0.02$ in our calculations.
\begin{figure}[h]
\includegraphics[width=0.42\textwidth]{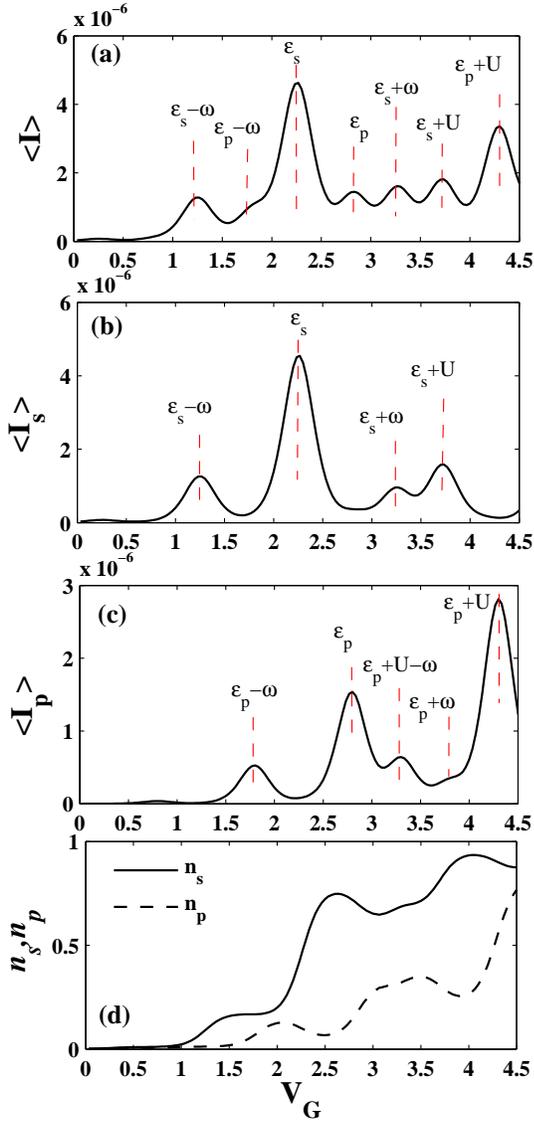}
\caption{\label{Fig2}  (a) The total average current $\langle
I\rangle$,  and (b,c) the average current  $\langle I_{s
(p)}\rangle$ for $s$($p$) energy level as a function of the gate
voltage $V_G$. (d) shows the average electron occupation number
$n_{s}$  and $n_{p}$. The intra-dot Coulomb interaction $U=1.5$,
$\Delta\varepsilon\equiv \varepsilon_p-\varepsilon_s=0.5$, and the
THz photon frequency $\omega=1$.
 \label{fig2}}
\end{figure}

\subsection{Analysis of
the experiment}

In the experiment \cite{Shibata}, the Coulomb charging energy
$E_{C}=15$ meV, which is comparable with the energy difference
between  $s$ and $p$ energy levels with $\Delta E=5$ meV. the THz
photon frequency is chosen to be larger than $\Delta E$ with
$hf_{\rm THz}=10.3$ meV. These parameters correspond to $U=1.5$,
$\Delta\varepsilon=0.5$, and $\omega=1$ ($\hbar=e=1$) in the
tunneling Hamiltonian (\ref{Hamiltonian}). In Fig. \ref{fig2}, we
plot the average currents and  electron occupation number as a
function of the gate voltage $V_G$. The main results are as follows.

First, in addition to the Coulomb blockade oscillation peaks
$\varepsilon_s$ and $\varepsilon_p+U$ as indicated in the
experiment, there also exists $\varepsilon_s+U$ peak, see Fig.
\ref{fig2}(a-b). In Fig. \ref{fig2}(d), we show that  $p$ energy
level has  a probability to be occupied. In this case, when an
electron tries to transit through the  $s$ level, it will be
accompanied by a Coulomb repulsion $U$. Therefore, for
$V_G=\varepsilon_s+U$, a resonant tunneling occurs.

Secondly, the photon-assisted side peaks at $\varepsilon_s\pm\omega$
can be observed, with the right side peak $\varepsilon_s+\omega$ in
$\langle I_{s}\rangle$ reduced slightly due to the competition with
the nearby $\varepsilon_s+U$ peak (Fig. \ref{fig2}(b)). However, in
the total average current $\langle I\rangle$, the
$\varepsilon_s+\omega$ peak coincides with the Coulomb blockade
oscillation PAT $\varepsilon_p+U-\omega$ in Fig. \ref{fig2}(c),
leading to the enhancement of this peak.

Thirdly, without THz irradiation, because of the Coulomb blockade,
the  $\varepsilon_p$ peak is strongly suppressed. Whereas under the
THz irradiation, an electron in  $s$  level can be excited into
leads, which reduces the Coulomb repulsion of  $p$ level and results
in the subsequent tunneling of electrons from leads through  $p$
level, see the PIER peak of $\varepsilon_p$ in Fig. \ref{fig2}(a,c).

Finally,  there exist two side peaks at $\varepsilon_p\pm\omega$ in
Fig. \ref{fig2}(c). For $V_G=\varepsilon_p-\omega$, both energy
levels are above the chemical potential of leads, an electron can
tunnel through $p$ level with the help of THz irradiation and one
can observe the PAT peak in the total average current (Fig.
\ref{fig2}(a)). Whereas for $V_G=\varepsilon_p+\omega$, the two
energy levels are below the chemical potential of leads. In this
case, the energy separation between $s$ level and leads is
$\omega+\Delta\varepsilon$, which is larger than the THz photon
frequency $\omega$. So, it is hard to excite an electron in $s$
level into leads, and results in a suppression of this side peak.

We now compare the above results with the InAs QD experiment. We
show that, the presented peaks of our model analysis
$\varepsilon_{s}$, $\varepsilon_p + U$, $\varepsilon_s \pm\omega$,
$\varepsilon_{p}$ agree well with $E_0$, $E_1$, $E_0\pm hf_{\rm
THz}$, and the PIER of $p$ level in Fig. 3 of the reference
\cite{Shibata}.  We find the side peak $\varepsilon_p-\omega$
induced by the THz irradiation in our  model analysis. This peak can
be identified in the original experimental data, but was not
addressed in  the reference \cite{Shibata}. In addition, beyond the
Coulomb blockade oscillation peak $\varepsilon_p + U$, we also find
there exists the $\varepsilon_s + U$ peak. This peak seems not
readily discriminated from $E_1$ ($E_1\equiv\varepsilon_p + U$) peak
in the experiment. In the future, one may expect to identify both
the peaks $\varepsilon_{s,p} + U$ by increasing the separation
between the energies $\varepsilon_s$ and $\varepsilon_p$.

We then compare our results with the MWF QD \cite{Oosterkamp,Sun}.
We see that beyond the PAT resonances of the two levels in MWF QD,
the Coulomb blockade oscillation peaks of $\varepsilon_{s,p}+U$
appear in the InAs QD. Moreover, we find that the photon-assisted
tunnelings of these Coulomb blockade oscillation peaks occur, which
also contribute to the side peaks of the $s$ level, see below for
detailed discussions.
\begin{figure}[h]
\includegraphics[width=0.39\textwidth]{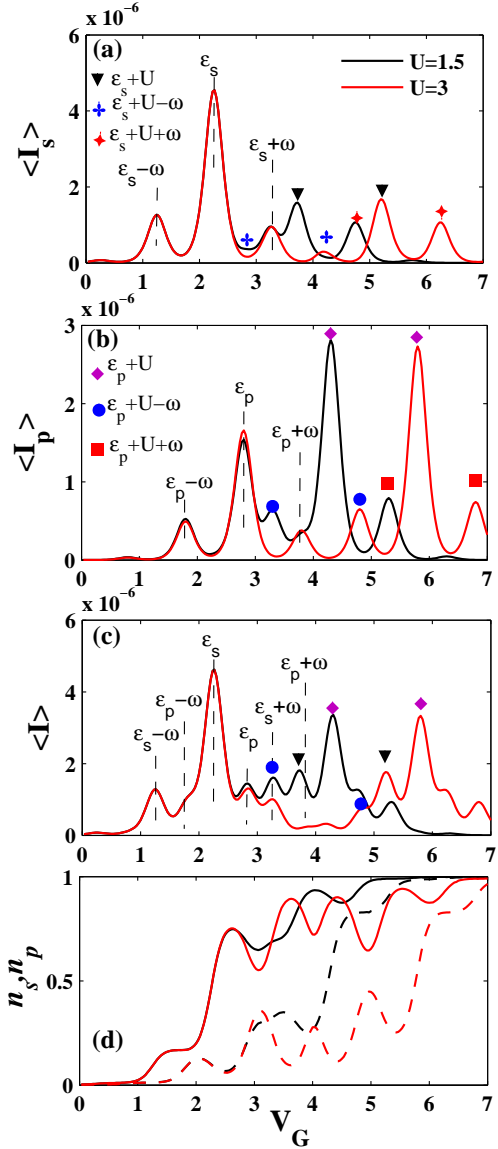}
\caption{\label{Fig3} (color online) (a,b) The average current
$\langle I_{s (p)}\rangle$ for $s$($p$) energy level, and (c) the
total average current $\langle I\rangle$ as a function of the gate
voltage $V_G$ for $\Delta\varepsilon<\omega<U$. (d) shows the
average electron occupation number $n_{s}$ (solid lines) and $n_{p}$
(dashed lines). \label{fig3}}
\end{figure}
\begin{figure}[h]
\includegraphics[width=0.39\textwidth]{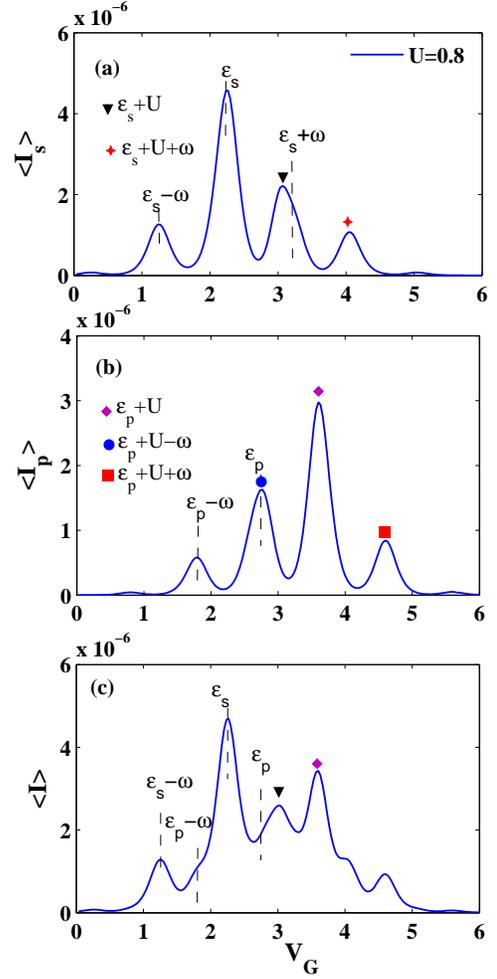}
\caption{\label{Fig4} (color online) (a,b) The average current
$\langle I_{s (p)}\rangle$ for $s$($p$) energy level, and (c) the
total average current $\langle I\rangle$ as a function of the gate
voltage $V_G$ for $\Delta\varepsilon<U<\omega$. \label{fig4}}
\end{figure}

\subsection{Coulomb interaction effects on the resonant tunnelings}
Now, we explore systematically the Coulomb interaction effects on
the resonant tunnelings. We first consider
$\Delta\varepsilon<\omega<U$. Fig. \ref{fig3} shows the results of
the average currents for $U=1.5$ and $3$. We see that, while the
main resonance $\varepsilon_s$, the PIER of $\varepsilon_p$, and the
side peaks $\varepsilon_{s,p}\pm\omega$ are not affected by
increasing $U$, the Coulomb interaction involved resonances, like
the peaks $\varepsilon_{s,p}+U$, and related PATs
$\varepsilon_{s,p}+U\pm\omega$ shift to higher gate voltage, but the
strengthen remains almost unchanged. Significantly, the Coulomb
blockade oscillation PATs $\varepsilon_s+U\pm\omega$ are asymmetric
with the peak $\varepsilon_s+U-\omega$ being largely suppressed
(Fig. \ref{fig3}(a)). This is because the occupation number $n_p$
for $V_G=\varepsilon_s+U-\omega$ is much smaller compared with
$V_G=\varepsilon_s+U+\omega$ (see $U=3$ in Fig. \ref{fig3}(d) for
example),  which makes an electron have little probability to
transit into the energy level $\varepsilon_s+U$  and thus reduces
the PAT. On the other hand, the Coulomb blockade oscillation PATs
$\varepsilon_p+U\pm\omega$ are quite symmetric, as shown in Fig.
\ref{fig3}(b).

In Fig. \ref{fig3}(c), we plot the total average current $\langle
I\rangle$ versus the gate voltage $V_G$. The tunneling processes can
be divided into three regimes, see $U=3$ for example. For the low
gate voltage $V_G\leq\varepsilon_s+\omega$, we can observe the main
resonance $\varepsilon_s$, the PIER of $\varepsilon_p$, and the
related side peaks $\varepsilon_{s}\pm\omega$ or
$\varepsilon_{p}-\omega$. For the high gate voltage
$V_G\geq\varepsilon_s+U$, the peaks $\varepsilon_{s,p}+U$  become
the dominant tunneling processes. Whereas for $V_G$ between above
two regimes, one enters into the Coulomb blockade regime, where the
finite total average current arises from the side peak
$\varepsilon_{p}+\omega$ and Coulomb blockade oscillation PATs like
$\varepsilon_{s,p}+U-\omega$.

Finally, we discuss the Coulomb interaction effects for
$\Delta\varepsilon<U<\omega$. In this case, the  peaks
$\varepsilon_{s}+U$ (Fig. \ref {fig4}(a)) and $\varepsilon_{p}+U$
(Fig. \ref {fig4}(b)) now move to the low gate voltage regime and
merge with the side peak $\varepsilon_s+\omega$, making this side
peak hard to be distinguished. Significantly, although one can see
the PIER of $\varepsilon_p$ in the average current $\langle
I_{p}\rangle$ (Fig. \ref {fig4}(b)), because $\varepsilon_{s}+U$ now
dominates the tunneling process and lies very close to $p$ level,
the PIER of $\varepsilon_{p}$ can not be directly identified, see
Fig. \ref {fig4}(c). Furthermore, while the main resonance
$\varepsilon_s$, and the side peaks $\varepsilon_{s,p}-\omega$ are
not affected, the Coulomb blockade oscillation PATs
$\varepsilon_{s,p}+U-\omega$ become featureless.

\section{Conclusion}
In conclusion, we have explored the THz photon-assisted tunneling
through a two-level InAs QD. Because the Coulomb interaction is of
the same order as the THz energy difference, the finite  Coulomb
interaction plays important roles on the tunneling processes. We
demonstrate that, the Coulomb blockade oscillation and PIER of $p$
level can be clearly observed. Beyond these results, we find new
Coulomb blockade oscillation and PAT peaks, which may be identified
in further experiment. In particular, we find that, to observe the
interesting photon-induced excited state resonance of $p$ level, the
Coulomb interaction should be larger than  THz photon frequency. We
believe these features are of practical importance for future THz
devices, for example, developing a highly sensitive and
frequency-tunable THz detector.

This work is supported by  NCET, NSFC under grants Nos. 11174168,
11074175, NKBRSFC under grants No.
 2011CB606405, and National Special Fund for the
Development of Major Research Equipment and Instruments under grants
No. 2011YQ13001805.

\end{document}